\documentclass[%
 reprint,
 amsmath,amssymb,
 aps,
]{revtex4-1}

\usepackage{graphicx}
\usepackage{dcolumn}
\usepackage{bm}
\usepackage{color}
\usepackage{romannum}
\usepackage{comment}

\usepackage[normalem]{ulem}
\usepackage{censor}
\usepackage{hyperref}
\graphicspath{./figures/}
\begin{document}
\title{Reversible-to-irreversible transition of colloidal polycrystals under cyclic athermal quasistatic deformation}
\author{Khushika$^{1}$, Lasse Laurson$^{2}$, Pritam Kumar Jana$^{1}$}
\email{pritam.jana@pilani.bits-pilani.ac.in}
\affiliation{$^1$Department of Chemistry, Birla Institute of Technology and Science, Pilani, Pilani Campus, Rajasthan 333031, India}
\affiliation{$^2$Computational Physics Laboratory, Tampere University, P.O. Box 692, FI-33014 Tampere, Finland}
\begin{abstract}
Cyclic loading on granular packings and amorphous media exhibits a transition from reversible elastic behavior to irreversible plasticity. The present study compares the irreversibility transition and microscopic details of colloidal polycrystals under oscillatory tensile-compressive and shear strain. Under both modes, the systems exhibit a reversible to irreversible transition. However, the strain amplitude at which the transition is observed is larger in the shear strain than in the tensile-compressive mode. The threshold strain amplitude is confirmed by analyzing the dynamical properties, such as mobility and atomic strain (von-Mises shear strain and the volumetric strain). The structural changes are quantified using a hexatic order parameter. Under both modes of deformation, dislocations and grain boundaries in polycrystals disappear, and monocrystals are formed. We also recognize the dislocation motion through grains. The key difference is that strain accumulates diagonally in oscillatory tensile-compressive deformation, whereas, in shear deformation, strain accumulation is along the $x$ or $y$ axis.
\end{abstract}

\pacs{Valid PACS appear here}
\maketitle


\section{\label{sec:level1}Introduction}
A large number of periodically driven soft materials, including disordered 
granular media \cite{Slotterback,Mobius,Royer}, colloidal suspensions \cite{Corte,tjhung2015hyperuniform,Besseling,haw1998direct}, colloidal gels \cite{Smith}, exhibit an intriguing transition from reversible, elastic behaviour to irreversible, plastic deformation as the applied strain amplitude overcomes a threshold value \cite {pine2005chaos,menon2009universality,leishangthemyielding,priezjevyielding,Keim2013,perez2018well, reichhardt2023reversible}. It is important to understand the origin of irreversibility from reversible microscopic dynamics because it may shed light on the nature of yielding in those systems. For example, in a crystalline
system, yielding is mediated by
defects motion. For amorphous solids, it is considered that localized
rearrangements, known as shear transformations, are responsible for flow [17]. However, the identification of those
local events due to their disordered structure is challenging. In recent work, the connection between irreversibility transition with yielding of amorphous solid and jamming is established by investigating the response of soft-sphere assemblies to athermal cyclic-shear deformation over a variety of densities and amplitudes of deformation \cite{nagasawa2019classification,das2020unified}. The relation between yielding in crystals and glasses is still a subject of current investigations \cite{PhysRevE.98.062607}.

In between those
two classes of materials, crystals and amorphous solids, lie
polycrystals, where several crystalline regions are separated
by grain boundaries.  The grain boundaries in polycrystalline materials can control bulk properties such as electrical conductivity, yield strength, etc. For example, the yield strength of the materials can be improved by increasing the density of grain boundaries \cite{hansen2004hall}. This structure-property relationship is even more interesting in two-dimensional (2D) materials where grain boundaries can translate, create new, or annihilate entirely from the system under external forces \cite{lobmeyer2022grain}.\newline
Colloidal particles are considered scaled-up models of atoms. Colloidal suspensions are largely used systems to directly observe phenomena that would otherwise not be within experimental reach. They can be assembled into colloidal polycrystals, where ordered crystalline regions are separated by extended grain boundaries formed by dislocation arrays \cite{Ghofraniha,Louhichi,Keim,Tamborini}. The application of a cyclic deformation to colloidal polycrystals allows to follow at the same time dynamics of individual particles (“atoms”) and the large-scale response of the poly-crystalline texture \cite{Ghofraniha,Louhichi,Keim,Tamborini}. Plastic deformation of colloidal polycrystals has been studied in experiments and simulations \cite{Buttinoni,Tamborini,Meer}. 
In a previous study, where we applied oscillatory shear deformation on polycrystalline samples, a non-equilibrium phase transition mediated by the motion of defects and controlled by the strain amplitude is observed \cite{jana2017irreversibility}. Experiments of plastic flow have also been performed under uniaxial stress in metallurgy, amorphous metals \cite{schuh2007mechanical}, and granular materials \cite{desrues1985localization}. A detailed comparison of the motion of defects that leads to non-equilibrium phase transition in different modes of deformations may shed light on microscopic dynamics associated with the process. 
\newline
\begin{figure*}
    \centering
    \includegraphics[scale=0.35]{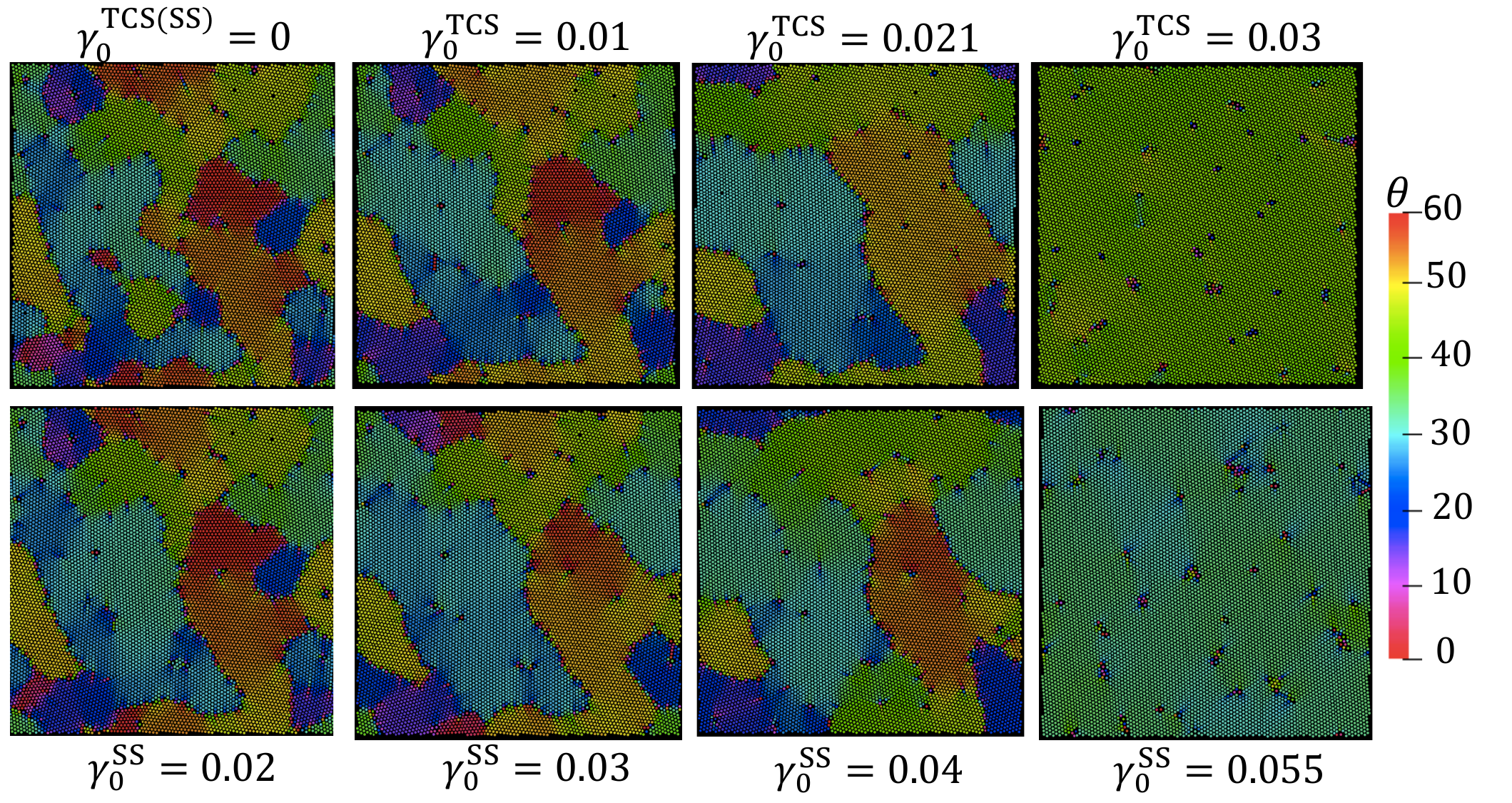} 
   \caption{Local grain orientations are shown after the completion of 200 cycles for different strain amplitudes. $\gamma_0^\mathrm{TCS(SS)}=0$ denotes the initial configuration of the sample, i.e., without any deformation. The top panel is for tensile-compressive strain, and the bottom panel is for shear strain.
   }
\label{Fig:Fig1}
\end{figure*}
In the present study, we explore the irreversibility transition in 2D-polycrystalline samples under oscillatory tension-compression and shear deformation by employing the athermal quasistatic method. The results reveal that under both deformation modes, the system exhibits a reversible to irreversible transition. The threshold strain amplitude after which irreversibility is observed is larger in the case of the oscillatory shear deformation. We have confirmed it by analyzing dynamical properties, such as mobility and atomic strain, and structural behavior, such as hexatic order parameters. Under both modes of deformation, we observe the disappearance of dislocations and grain boundaries and the formation of monocrystals with some defects because of the particles with larger sizes. The mechanism of monocrystal formation is also identified where a key difference is that in the case of oscillatory tensile-compressive loading, the strain accumulates diagonally, whereas, in oscillatory shear strain, the strain accumulation is along the $x$- or $y$-axis.\newline 
This paper is organized as follows: in Sec. \Romannum{2} we describe the model and simulation details, whereas the analysis, results, and discussions are presented in Sec. \Romannum{3}. Finally, the concluding remarks are made in Sec. \Romannum{4}.
\section{\label{sec:level1}Model and simulation details}
We consider a two-dimensional system with two types of particles, large ($l$) and small ($s$). They are interacting via a pairwise Lennard-Jones potential 
\begin{equation}
V_{\alpha\beta}(r)=4\epsilon_{\alpha\beta}\left[\left(\frac{\sigma_{\alpha\beta}}{r}\right)^{12}-\left(\frac{\sigma_{\alpha\beta}}{r}\right)^{6}\right]-V_{\alpha\beta}(r_c)
\end{equation}
where $r$ is the distance between the particle of type $\alpha$ and one of type $\beta$ ($\alpha,\beta=l,s$). The function $V_{\alpha,\beta}(r)=0$ when $r>r_c^{\alpha\beta}$ a cut-off distance and $V_{\alpha\beta}(r_c)$ ensures that $V_{\alpha\beta}(r)$ is continuous at $r=r_c^{\alpha\beta}$. The parameters of the potential are chosen as follows: $\sigma\equiv\sigma_s=1.0$, $\sigma_l=1.4\sigma$ and $\sigma_{ls}=1.2\sigma$, where $\sigma_{\alpha,\beta}$ is the finite distance at which inter-particle ($\alpha, \beta$) potential becomes zero (are the particle diameter) and $\varepsilon\equiv  \varepsilon_{ll}=1.0$, $\varepsilon_{ss}=0.5\varepsilon$, $\varepsilon_{ls}=1.5\varepsilon$ are the energy parameters. The cut-off distance ${r_c}^{\alpha\beta}$ is fixed to $3.0\sigma_{\alpha\beta}$. The typical numbers of the particles used in the system are $N_l=50$ and $N_s=10000$ with $m=1$ for both. The typical length of a 2D simulation box is $L=100\sigma$. The unit of the time is set $\tau=\sigma\sqrt{m/\varepsilon}$.\\
First, we prepare the polycrystalline sample as shown in Fig. \ref{Fig:Fig1}. A polycrystalline structure is made as follows: the binary mixture is heated at the temperature of $2.0 \varepsilon/k_B$ for $50\tau$, and then the temperature is reduced to $0.001 \varepsilon/k_B$ rapidly by a span of $5\tau$ and then equilibrated for another $5\tau$ at the same temperature. The equations of motion are solved numerically using the time-reversible measure-preserving Verlet and reversible reference system propagator algorithms integration
scheme with a time step $\Delta t=0.005\tau$. The temperature of the system is controlled by connecting to the Nosé-Hoover thermostat. Finally, the reminiscent kinetic energy is drained out by performing NVE simulations with the viscous drag $1.0 \varepsilon\tau/\sigma^2$ by a time span of $250\tau$ as done in Ref. \cite{Negri}.  \newline 
The dynamics under the deformation is athermal quasistatic \cite{regev2013onset,parmar2019strain}. Lees-Edwards periodic boundary conditions are employed. In each deformation step, a small strain increment $10^{-3}$ is followed by energy minimization using the conjugate gradient method. The strain is applied in a periodic manner: First, positive strain steps are applied. When a maximal predecided strain $\gamma_0$ is reached, the strain is reversed by applying strain steps in the opposite direction. This proceeds until the strain reaches the negative value of the maximal strain $-\gamma_0$. At this point, the strain steps are reversed until the system returns to zero strain, completing the cycle ($0\rightarrow \gamma_\mathrm{0} \rightarrow 0 \rightarrow -\gamma_\mathrm{0}\rightarrow 0$). The cycle is then repeated for $n_\mathrm{max}=200$ times. The position of all atoms was saved at the end of every cycle. We have applied the deformation on 5 to 10 samples prepared independently. \\The deformation of the sample is carried out in two different ways. In one case, we apply volume-preserving periodic shear strain (SS) along the $xy$ plane by incrementing the strain via the coordinate transformation of $x^{'}=x+yd\gamma_{xy}$, and $y^{'}=y$.  In the other case, we apply the deformation along the $x$ direction. Specifically, when there is an elongation of the sample along the $x$-directions, to preserve the volume, there is compression along the $y$ direction, and during the $x$-axis compression, elongation along the $y$-direction is conducted, called periodic tensile-compressive strain (TCS). The volume-preserving cyclic deformation is a well-accepted approach employed to explore the yielding, irreversibility transition, etc., of colloidal suspensions, amorphous solids, and polycrystals \cite{leishangthemyielding,parmar2019strain,Regev,jana2017irreversibility,PhysRevE.98.062607}. On a passing note, in systems with high particle density, the dynamics are primarily influenced by particle-particle interactions rather than particle-solvent interactions \cite{van2014molecular}. Therefore, the solvent effect is negligible. When dealing with attractive colloidal particles, one common approach is to employ the square well potential \cite{zaccarelli2002confirmation}, Lennard-Jones (LJ) potential shifted by the diameter of the particle \cite{patel2005interactions, santos2013effective}. The simulations are performed using LAMMPS \cite{Plimpton}. 



\section{\label{sec:level1}Results and discussion}
\begin{figure*}
    \centering
    (a)\includegraphics[scale=0.45]{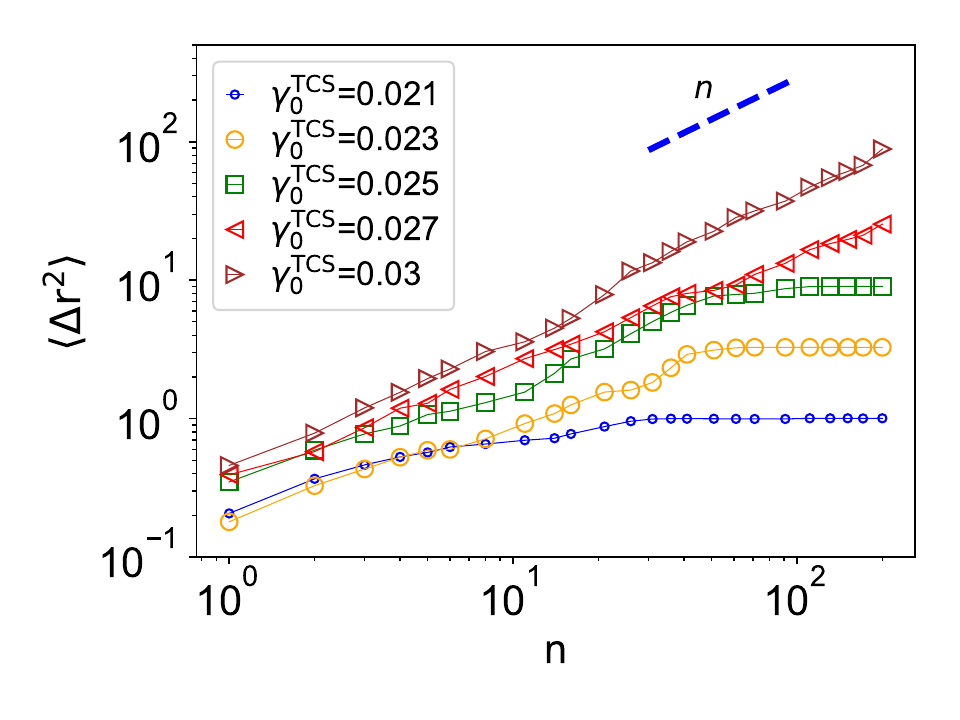}
    (b)\includegraphics[scale=0.45]{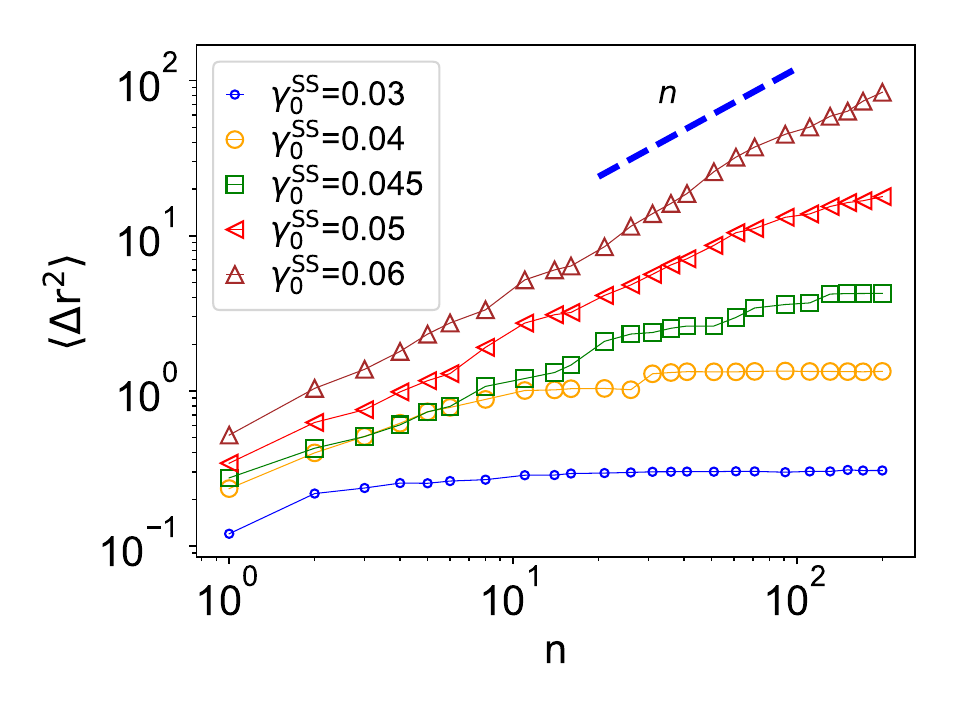}  
     (c)\includegraphics[scale=0.45]{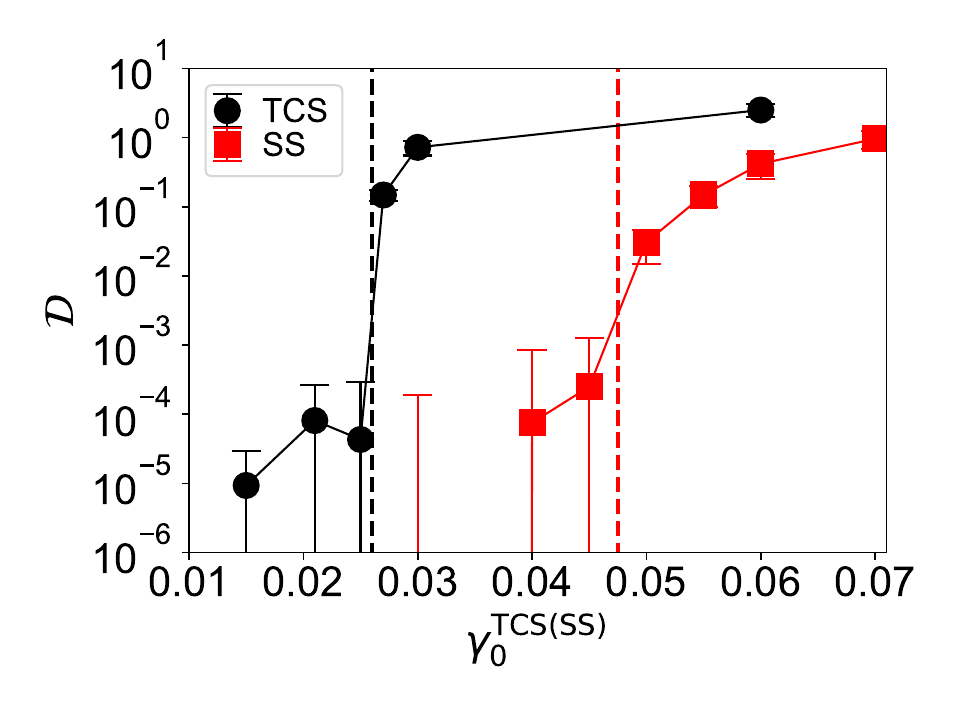}
     (d)\includegraphics[scale=0.45]{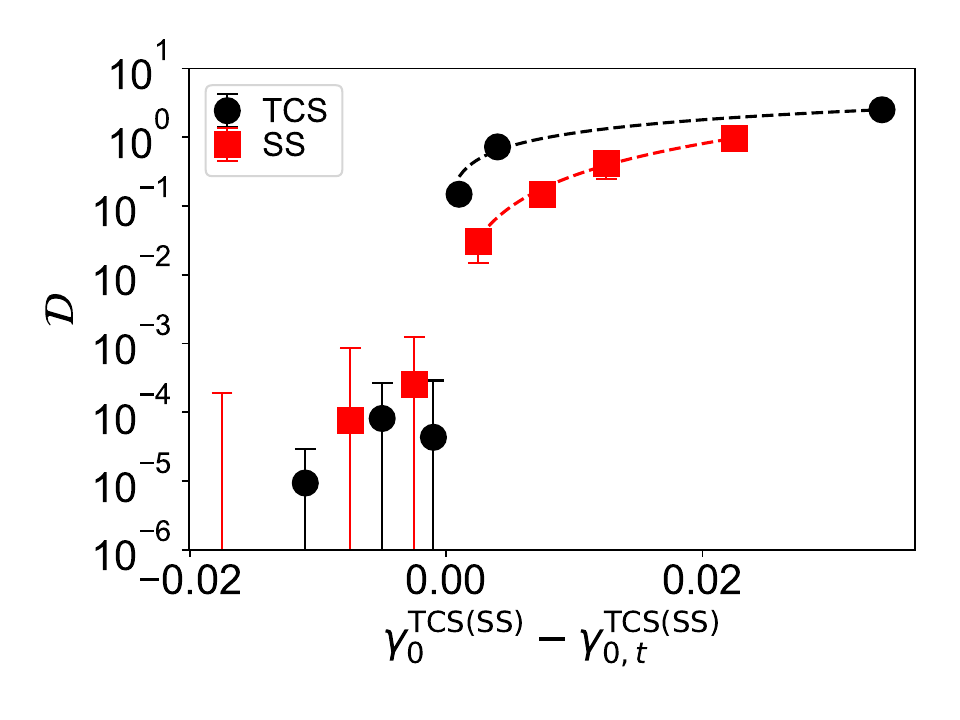}
   \caption{MSD as a function of the number of cycles, $n$, for (a) TCS and for (b) shear strain. (c) Local slope averaged over within last 50 cycles, $\mathcal{D}$, is shown as a function of strain amplitudes, for TCS and for shear strain. The dashed lines indicate the threshold strain amplitudes, $\gamma_{0,t}^\mathrm{TCS(SS)}$. (d) $\mathcal{D}$ is shown as a function $\gamma_{0}^\mathrm{TCS(SS)}-\gamma_{0,t}^\mathrm{TCS(SS)}$. The dashed lines indicate $\mathcal{D}\sim\left[\gamma_{0}^\mathrm{TCS(SS)}-\gamma_{0,t}^\mathrm{TCS(SS)}\right]^\alpha$ with $\alpha=0.64$ and $1.54$ for TCS and SS, respectively.}
    \label{Fig:Fig2}
\end{figure*}
\begin{figure*}
    \centering
    (a)\includegraphics[scale=0.4]{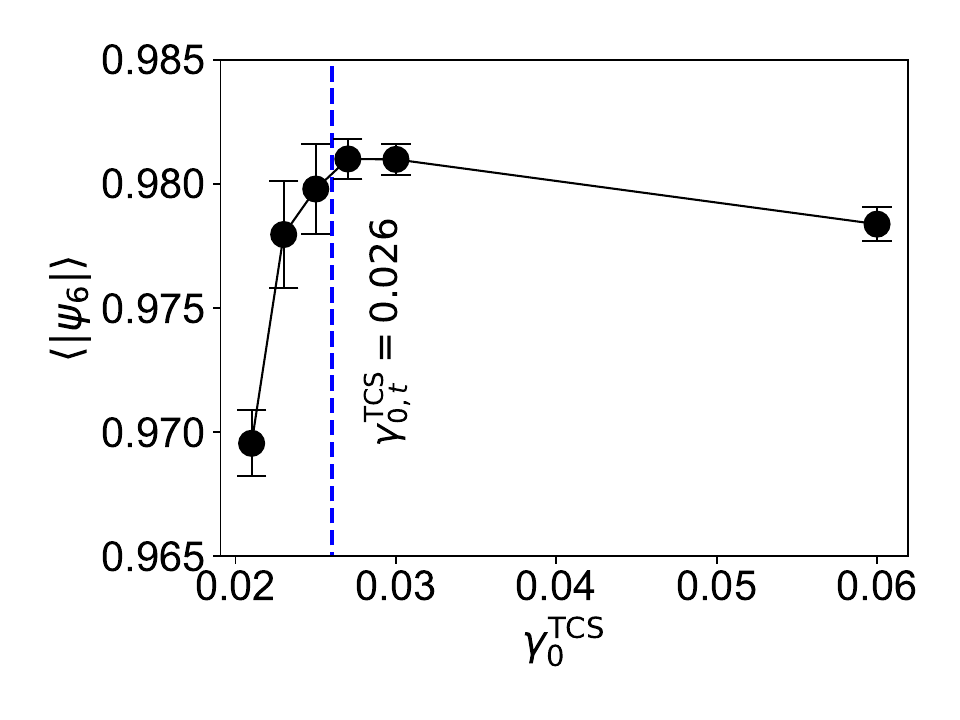}
    (b)\includegraphics[scale=0.4]{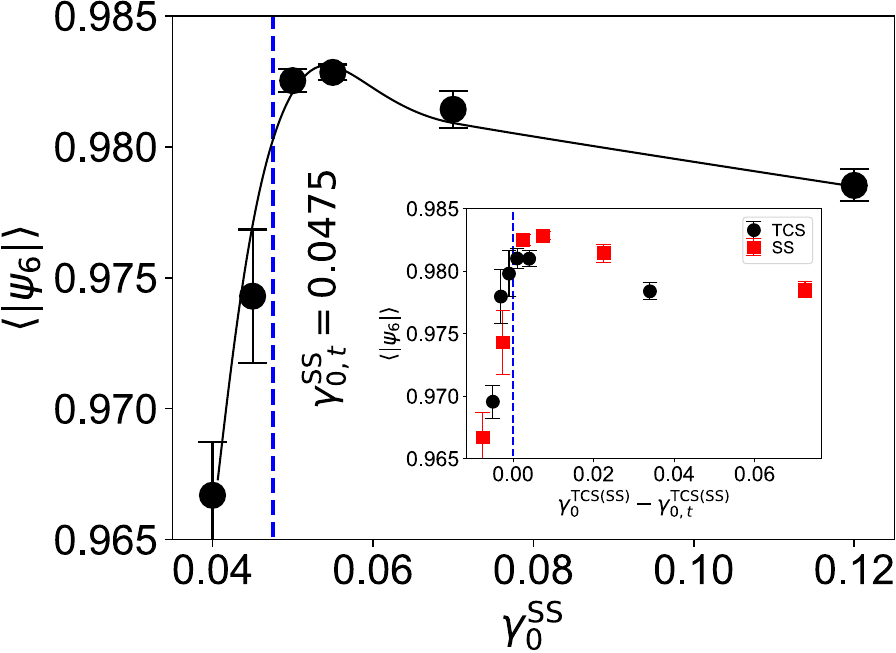}
    (c)\includegraphics[scale=0.4]{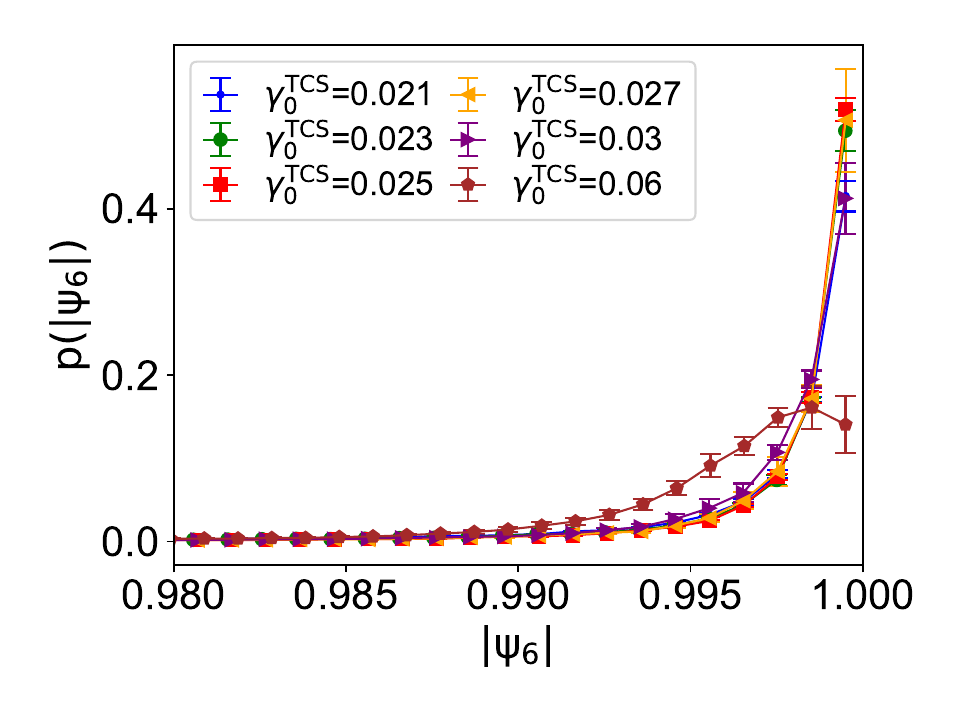}
    (d)\includegraphics[scale=0.4]{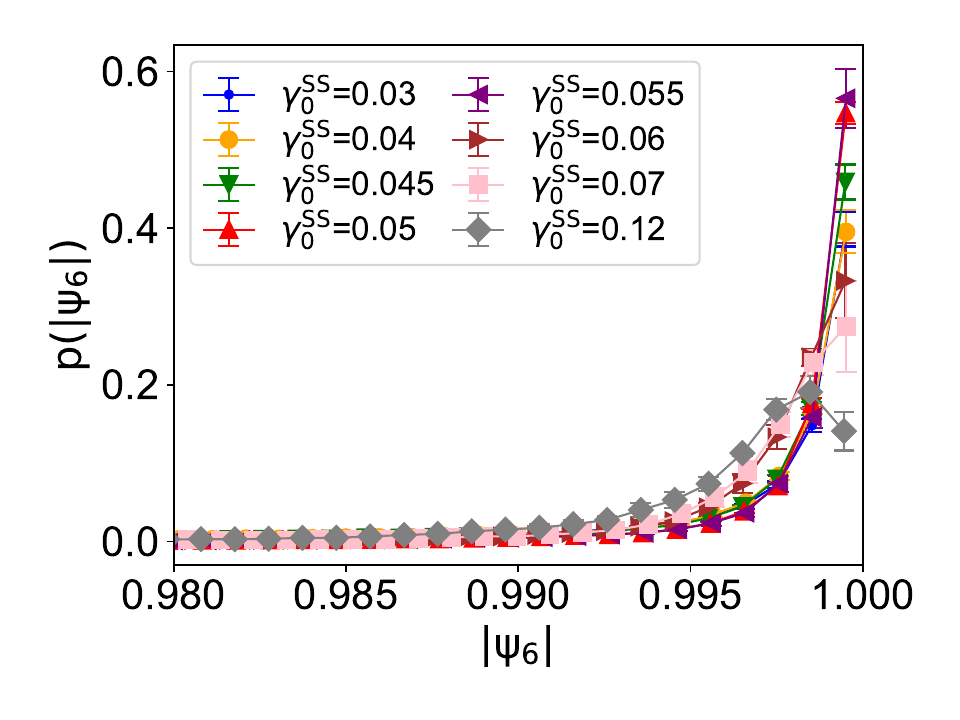}
   \caption{Average hexatic order parameter, $\langle|\psi_6|\rangle$, as a function of (a) tensile-compressive strain amplitude, $\gamma_0^\mathrm{TCS}$ and (b) shear strain amplitude, $\gamma_0^\mathrm{SS}$. The dashed blue line indicates the threshold value of irreversible transition. In inset we display $\langle|\psi_6|\rangle$ as a function of $\gamma_0^\mathrm{TCS(SS)}-\gamma_{0,t}^\mathrm{TCS(SS)}$.   
   Distribution of hexatic order parameter, $|\psi_6|$ under (c) tensile-compressive and (d) shear strain for different values of amplitudes. The solid lines are the guide to the eyes.}
    \label{Fig:Fig3}
\end{figure*}
\begin{figure*}
    \centering
        (a)\includegraphics[scale=0.42]{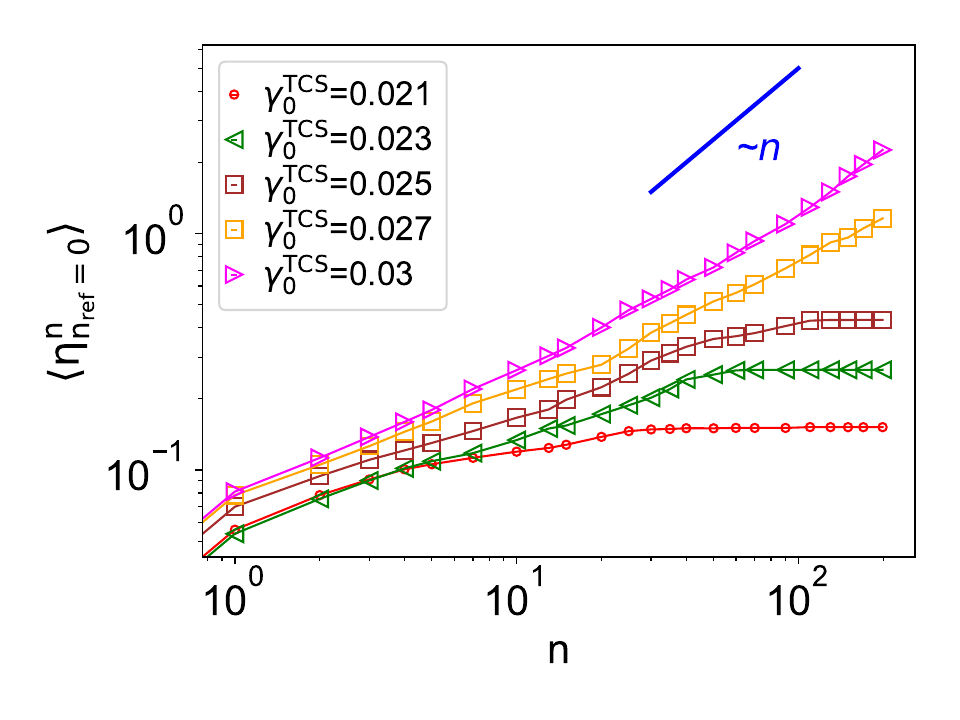} 
        (b)\includegraphics[scale=0.4]{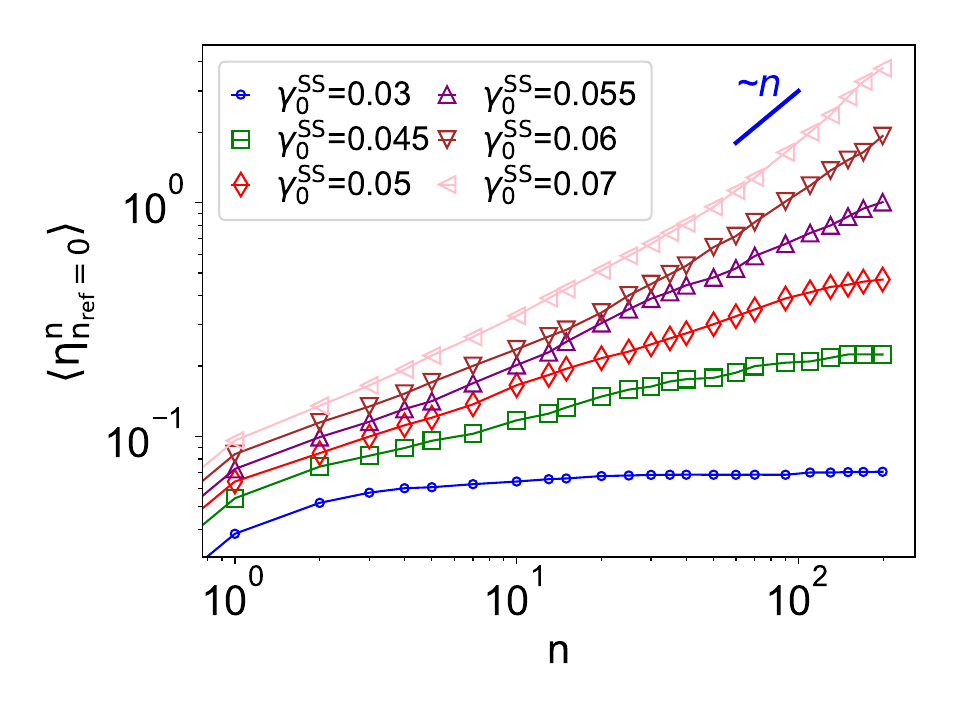}
        (c)\includegraphics[scale=0.4]{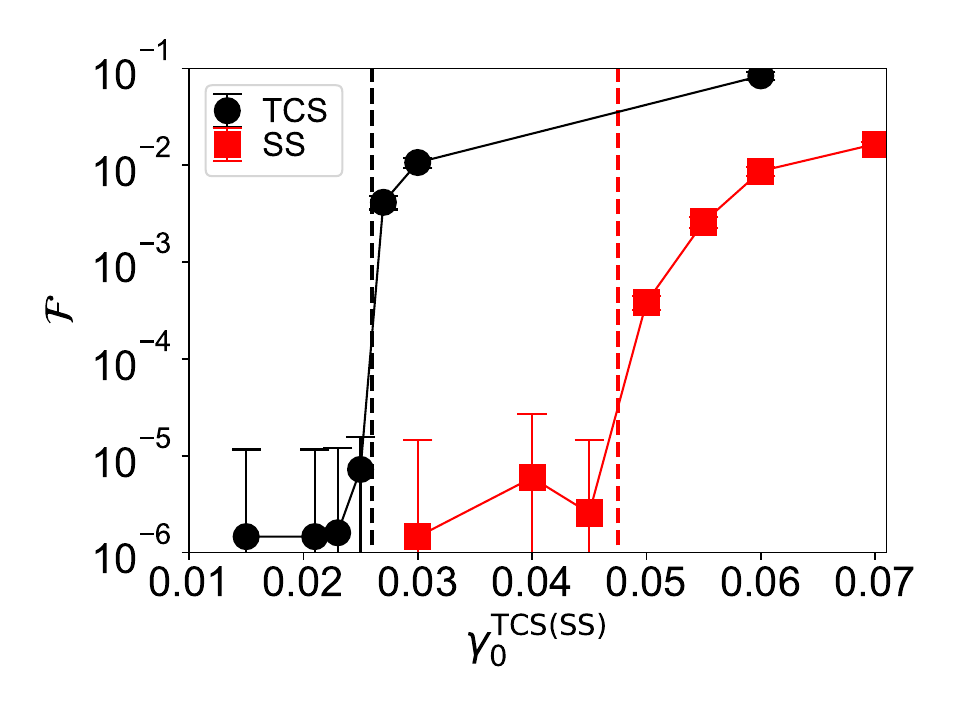}
        (d)\includegraphics[scale=0.4]{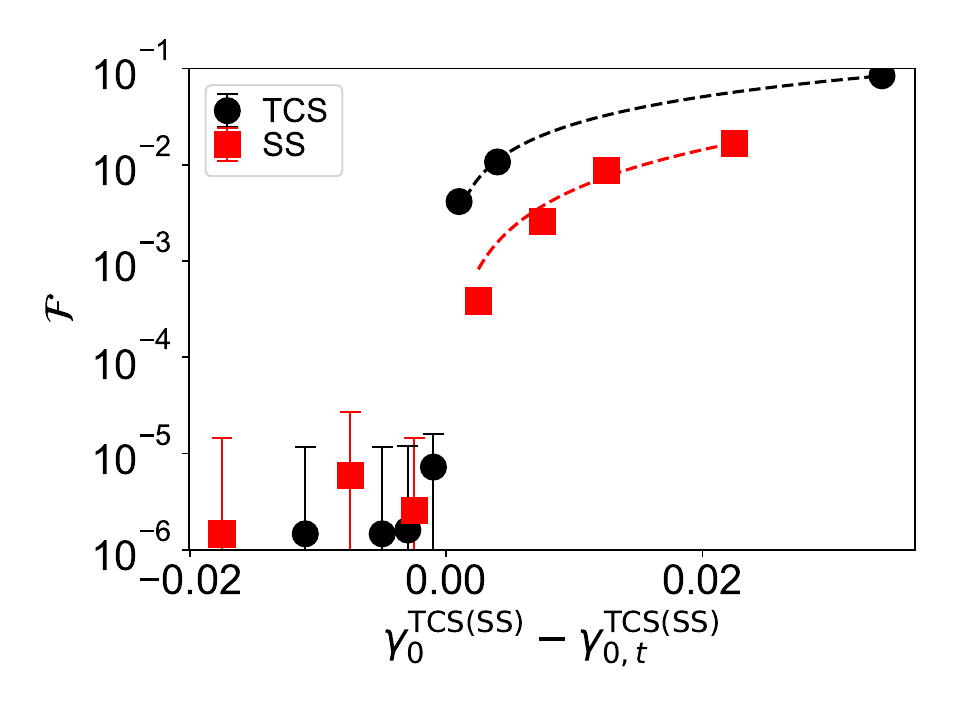}
   \caption{Average of von-Mises shear strain, $\langle\eta_{n_{\mathrm{ref}=0}}^n\rangle$, as a function of the number of cycles, $n$, for (a) TCS and for (b) shear strain. (c) Local slope averaged over within last 50 cycles, $\mathcal{F}$, is shown as a function of strain amplitudes for TC and shear strain. The dashed lines indicate the threshold strain amplitudes, $\gamma_{0,t}^\mathrm{TCS(SS)}$. (d) $\mathcal{F}$ is shown as a function of $\gamma_{0}^\mathrm{TCS(SS)}-\gamma_{0,t}^\mathrm{TCS(SS)}$. The dashed lines indicate $\mathcal{F}\sim \left[\gamma_{0}^\mathrm{TCS(SS)}-\gamma_{0,t}^\mathrm{TCS(SS)}\right]^{\beta}$ with $\beta=0.94$ and $1.38$ for TCS and SS, respectively. }
    \label{Fig:Fig4}
\end{figure*}
\begin{figure*}
    \centering
        a)\includegraphics[scale=0.4]{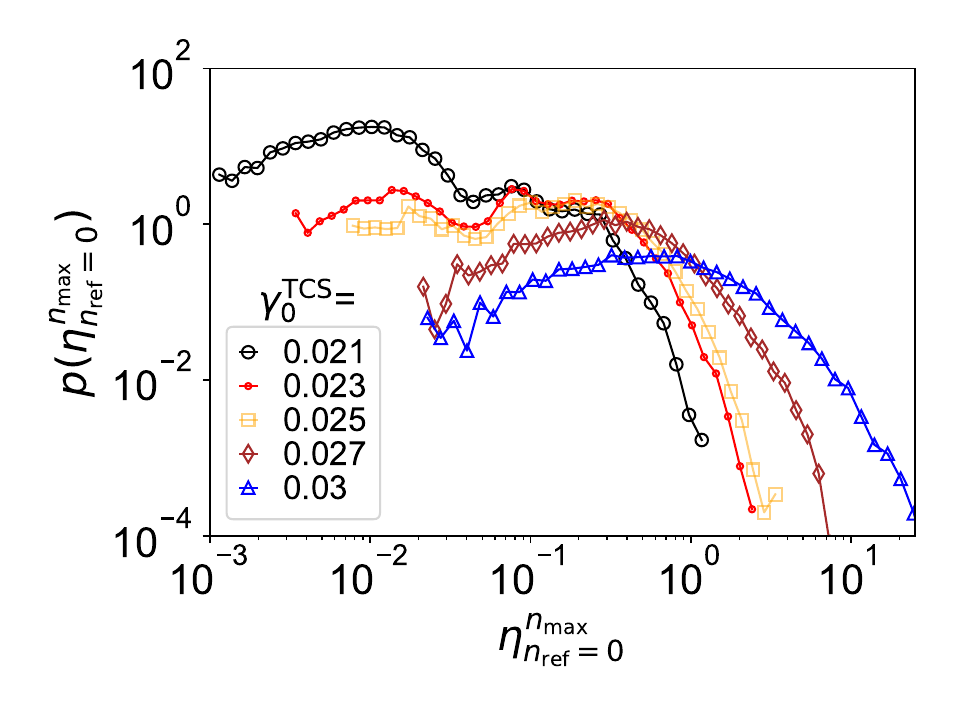}
        b)\includegraphics[scale=0.4]{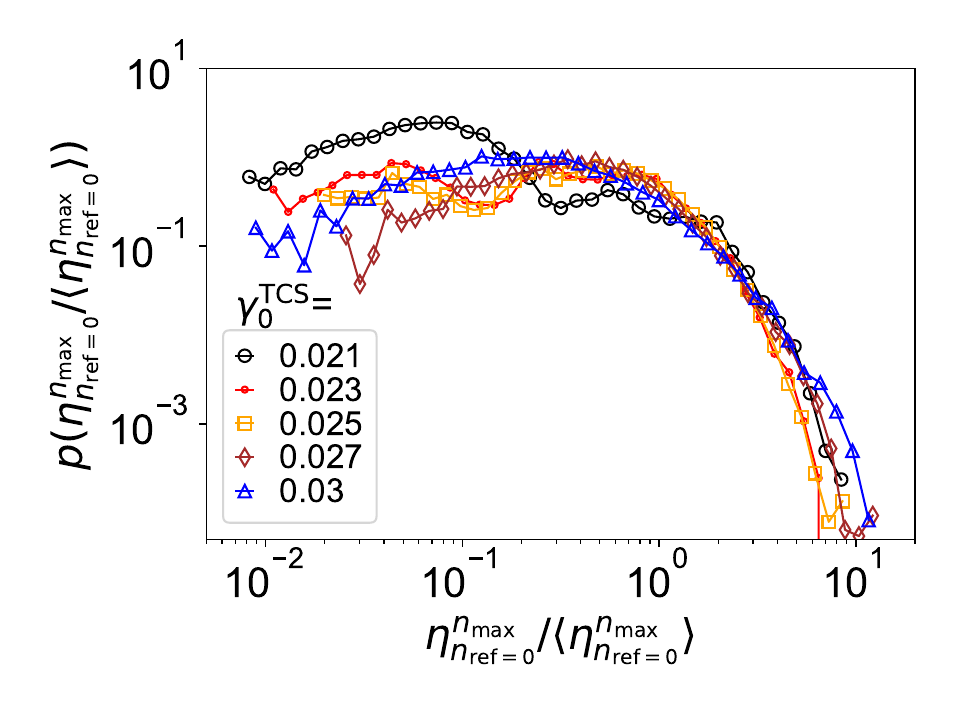}
        c)\includegraphics[scale=0.4]{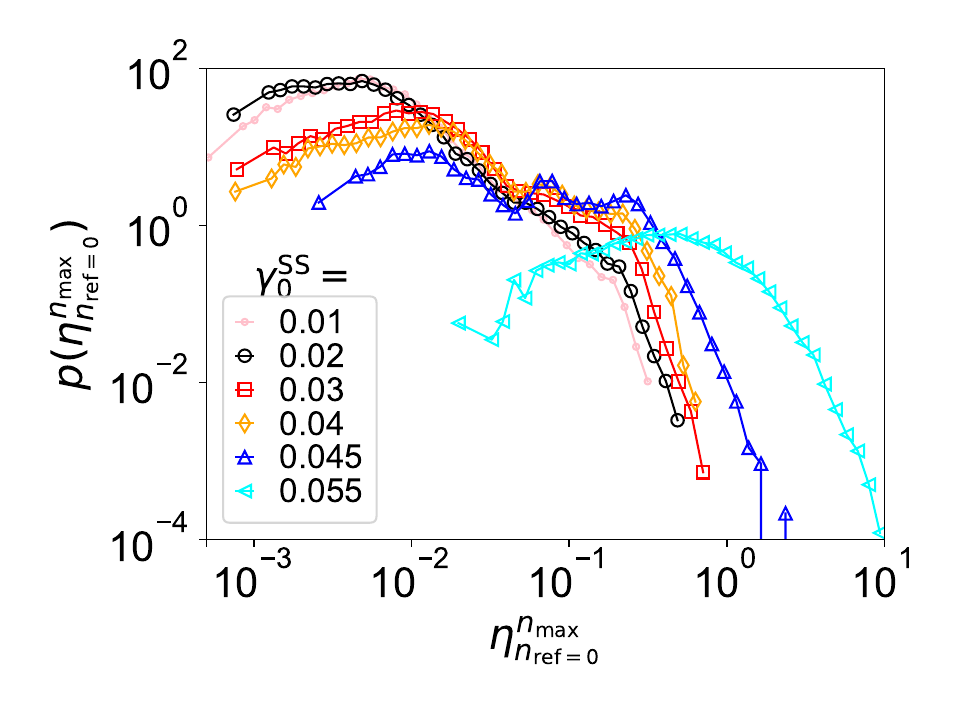}
        d)\includegraphics[scale=0.4]{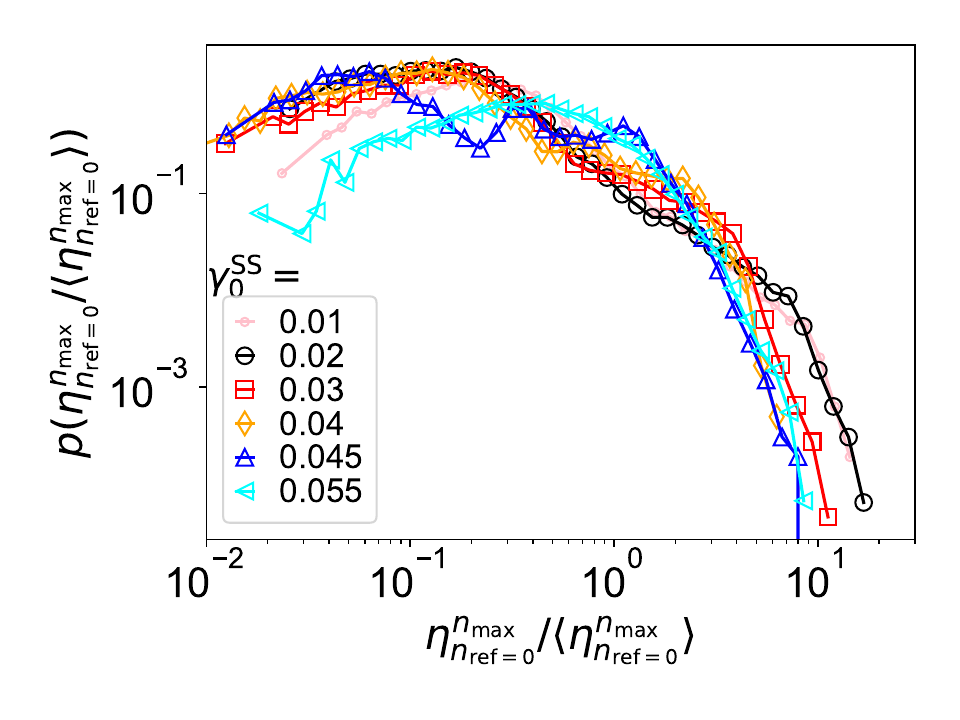}
   \caption{Distribution of von-Mises shear strain measured after completion of $n_\mathrm{max}(=200)$ cycles for the (a) TCS and (c) shear strain with respect to the initial configuration. The data are collapsed using average accumulated strain and shown in (b) and (d) for TC and shear strain, respectively. 
   }
    \label{Fig:Fig5}
\end{figure*}
\begin{figure*}
    \centering
        \includegraphics[scale=0.48]{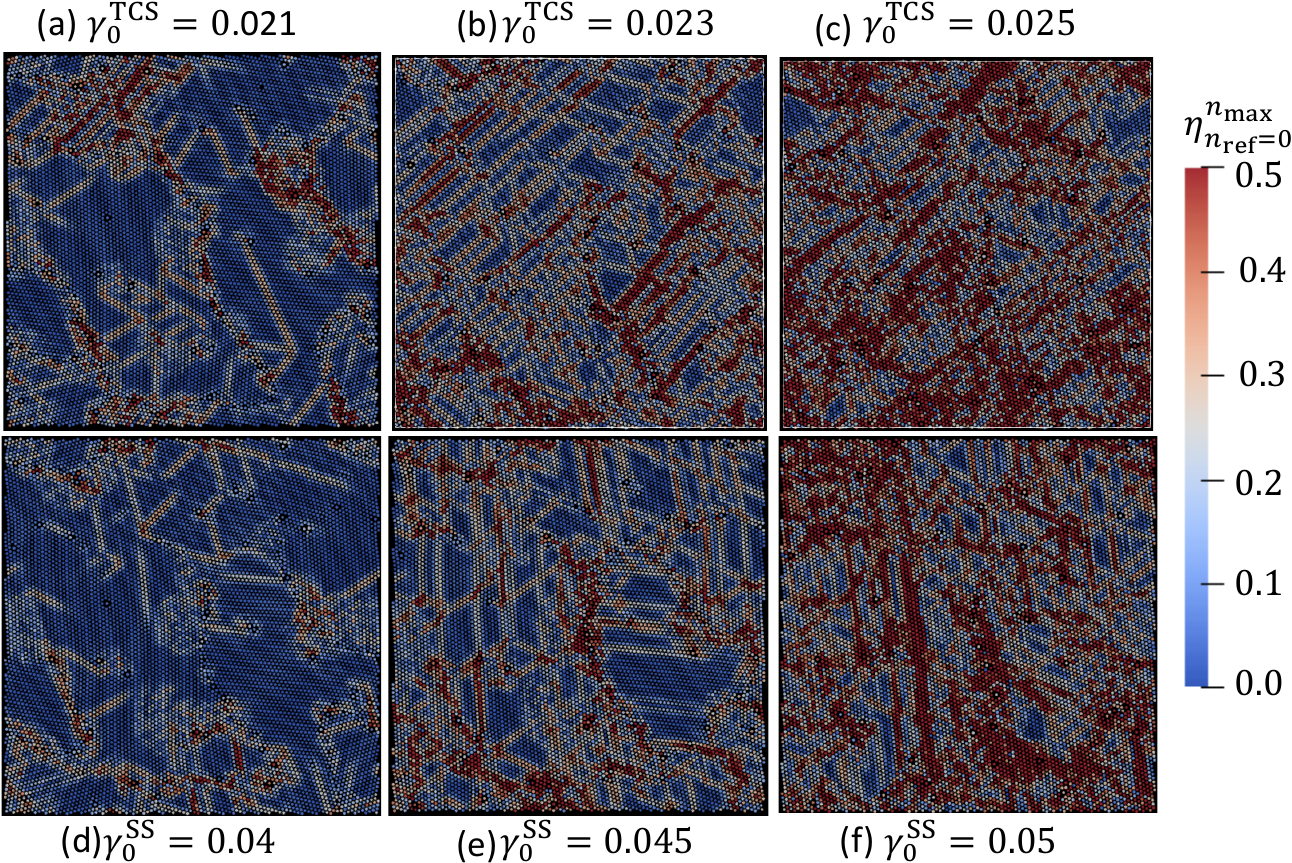}
   \caption{von-Mises shear strain map for individual particles after 200 cycles. It is measured with respect to the initial configurations. The top panel (a,b,c) corresponds to TCS, and the bottom panel (d,e,f) is for shear strain. The strain amplitude is specified. For the tensile-compressive loading, the strain accumulates diagonally, and for the shear loading, strain is accumulated along the $x$ or $y$ axis.
   }
    \label{Fig:Fig6}
\end{figure*}

\begin{figure*}
    \centering
(a)\includegraphics[scale=0.45]{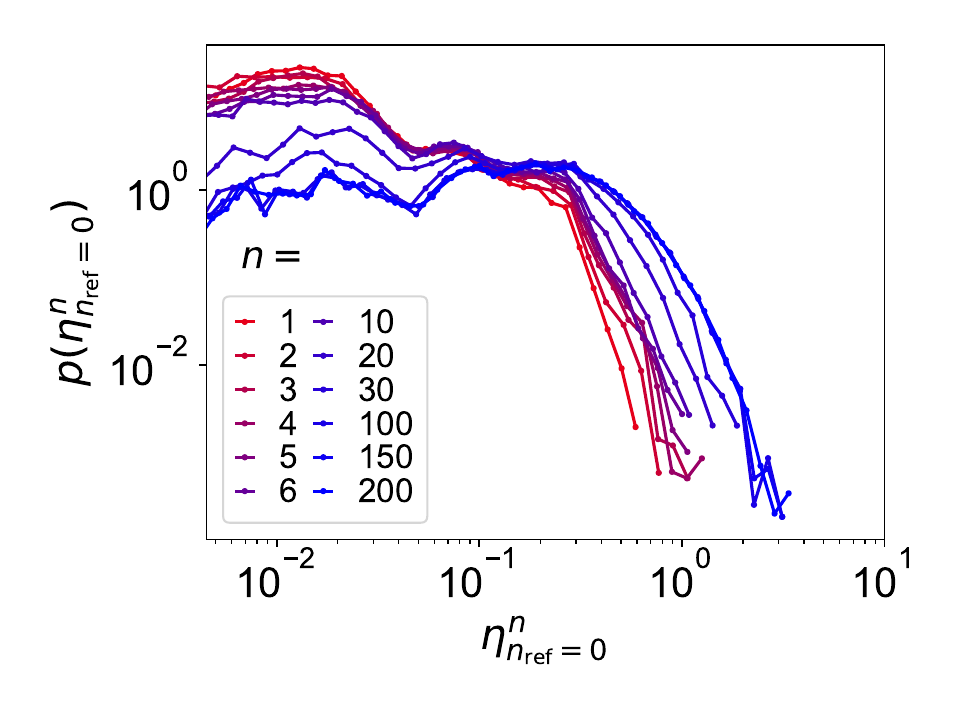}
(b)\includegraphics[scale=0.45]{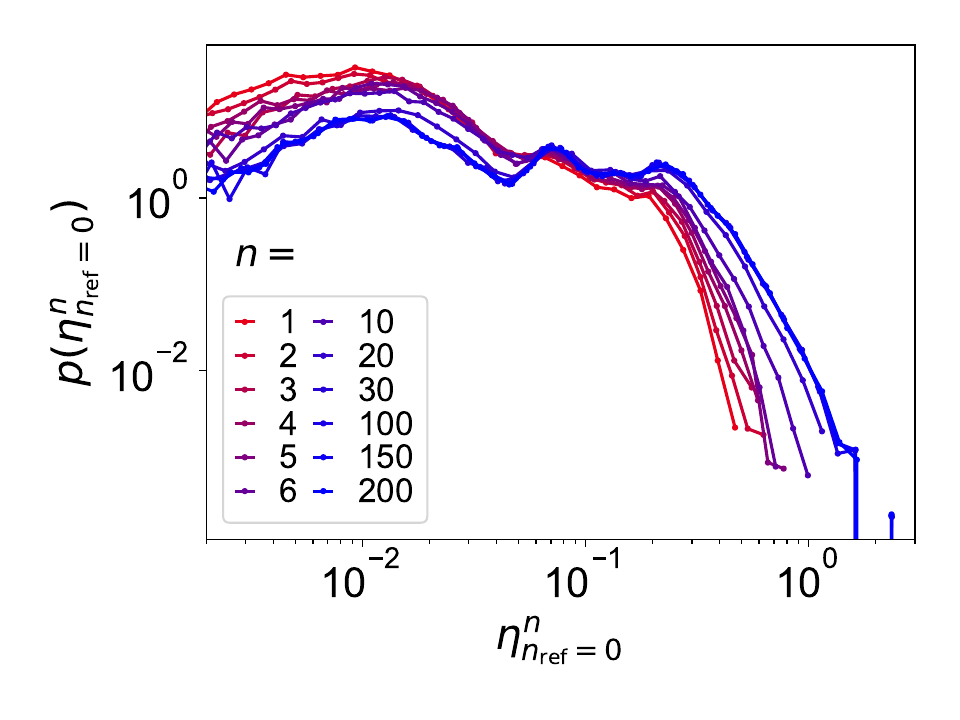}
\caption{von-Mises shear strain of individual particle is measured after $n$ cycles with respect to the initial configuration, i.e., $n_\mathrm{ref}=0$ and the distribution is displayed. (a) is for $\gamma_0^\mathrm{TCS}=0.025$ and (b) is for $\gamma_0^\mathrm{SS}=0.045$.}
    \label{Fig:Fig7}
\end{figure*}
In 2D triangular colloidal crystals, five- and seven-coordinated particles, known as positive (+1) and negatively charged disclinations ($-1$), exist as defects. The likelihood of an individual disclination occurring is low. However, combining five- and seven-coordinated particles with small additional energy creates edge dislocations. A regular array of dislocations then forms a grain boundary separating differently oriented crystals, as shown in Fig. \ref{Fig:Fig1}. A careful study shows that arrays of dislocations are mainly arranged along the grain boundaries and create a large difference in the orientation of the adjacent grains, which is consistent with Frank condition, $\rho \propto \mathrm{sin}\text{ } d\xi$, where $\rho$ is the line density (arrays of dislocations) along the grain boundary and $d\xi$ represents the difference in orientation of the adjacent grains separated by the grain boundaries.
In Fig. \ref{Fig:Fig1}, we display the grain orientation of a sample before and after the completion of 200 cycles with a range of values of strain amplitudes for both modes of deformation. The properties of polycrystalline samples are primarily influenced by the number of dislocations and impurities along the grain boundaries. We maintained a fixed density of the larger particles, which serve as impurities, and examined the disclination density across 10 samples. The resulting average is $0.061\pm0.001 \text{ }\sigma^{-2}$, which suggests a small sample-to-sample variation.
The top panel corresponds to the tensile-compressive strain except for the figure in the left corner, which displays a sample without any deformation, and the bottom panel is for shear strain. When the amplitudes of the deformation are $\gamma_0^{\mathrm{TCS}}=0.01, 0.021$ and $\gamma_0^{\mathrm{SS}}=0.02, 0.03, 0.04$, after a limit cycle, the sample responses reversibly. However, when $\gamma_0^{\mathrm{TCS}}=0.03$ and $\gamma_0^{\mathrm{SS}}=0.055$, a single crystal is seen in both modes of deformation. \\
{\textbf{Enhanced mobility.}} Depending on the strain amplitude, the local rearrangements of particles can be reversible after one or more cycles or irreversible, leading to chaotic dynamics and particle diffusion \cite{Regev}. To characterize particle motion under cyclic deformation, we compute mean square displacement (MSD), $\langle \Delta r^2 \rangle$, at the end of each cycle, $n$, considering the initial configuration ($n=0$) as a reference which is defined as follows
\begin{equation}
\langle\Delta r^2(n)\rangle = {1\over N}\left\langle \sum_{i=1}^N |{\bf{r}}_i(n)-{\bf{r}}_i(0)|^2\right\rangle.
\end{equation}
We plot the MSD as a function of the number of cycles for TCS and SS in Figs. \ref{Fig:Fig2}(a) and \ref{Fig:Fig2}(b), respectively. The results reveal that for a small strain amplitude, after a couple of cycles, particles exhibit a reversible behavior. However, with the increase of the strain amplitudes, it takes more and more cycles to eventually exhibit the reversible behavior. When $\gamma_0^\mathrm{TCS(SS)}$ is larger than the threshold value, we see that the system does not display any more reversibility within the explored time window, corresponding to irreversible, non-periodic particle trajectories and $\langle\Delta r^2\rangle$ there scales as $\langle\Delta r^2\rangle\sim n$. To quantify the mobility, we measure the local slope, $d\langle\Delta r^2\rangle/dn$, of $\langle\Delta r^2\rangle$ vs. $n$ plot and averaged over the steady states and denote it as diffusivity, $\mathcal{D}$ as we have done in our previous work \cite{jana2017irreversibility,PhysRevE.98.062607} and a similar approach is used in Ref. \cite{Fiocco,pine2005chaos,kawasakimacroscopic}. In Fig. \ref{Fig:Fig2}(c), we depict $\mathcal{D}$ as a function of strain amplitudes, $\gamma_0^\mathrm{TCS(SS)}$. The results show that for a low strain amplitude, the particle diffusivity vanishes in steady states, i.e., as expected $\mathcal{D}\approx0$. However, after a threshold value that appears as $\gamma_{0,t}^\mathrm{TCS}=0.026$ and $\gamma_{0,t}^\mathrm{SS}=0.0475$, the system exhibits a significant jump in the $\mathcal{D}$ and becomes diffusive within the simulation timescale. The response is similar in both modes of deformation. However, threshold strain amplitude in SS mode is larger by approximately a factor of 2 compared to the TCS mode. In TCS, the amplitude of the applied strain is associated with deformation along the x-direction. However, to maintain the volume of the simulation box, elongation in the x-direction necessitates a simultaneous compression along the y-direction. As a result, the material effectively undergoes deformation in both the x and y directions, a factor of 2 greater than the strain applied along the x-direction. In contrast, when dealing with volume-preserving shear deformation, no such phenomenon occurs. The geometric distinctions between these two deformation modes are linked to a twofold shift. In the previous work, we found the threshold value as $0.044$ for oscillatory shear deformation at temperature $T=0.001$ \cite{jana2017irreversibility,PhysRevE.98.062607}. Most probably, the difference appeared because of the different deformation protocols. Several experiments and simulations have attempted to quantify the nature of reversible to irreversible transition. Experiments with colloidal suspensions and numerical models show that the transition from reversible to irreversibility is a second-order non-equilibrium phase transition \cite{Corte}. While a numerical study with jammed solids shows the transition as non-equilibrium first-order \cite{kawasakimacroscopic}. However, an alternative explanation was also proposed \cite{PhysRevE.79.030101,regev2013onset}, which relies on the chaotic nature of trajectories in dynamical systems but not the phase transition as a requirement to explain the relatively sharp onset of irreversibility observed in the experiments.
In the present system, we recognize that near irreversibility transitions, $\mathcal{D}$ decreases by 1-2 orders of magnitude and that can be scaled algebraically, $\mathcal{D}\sim \left(\gamma_{0}^\mathrm{TCS(SS)}-\gamma_{0,t}^\mathrm{TCS(SS)}\right)^{\alpha}$ with $\alpha=0.64$ and $1.54$ for TCS and SS, respectively as $\gamma_{0,t}^\mathrm{TCS(SS)}$ is approached from above [see Fig. \ref{Fig:Fig2}(d)]. A recent study, where two-dimensional, amorphous solids under oscillatory shear are investigated, shows the diffusion coefficient above the transition follows power-law scaling with exponent $1.217>1$ \cite{regev2018critical}, while in the case of three-dimensional Lennard-Jones glass, the exponent appears as $0.61$ for low and $0.54$ for high temperature \cite{Fiocco}. 
\\
\textbf{Hexatic order parameter.} To further quantify how the merging of grains influences the local structure, we measure 2D local bond-orientational order parameter, $\psi_{6}$ for each particle $i$ and plot it in Fig. \ref{Fig:Fig3}. 2D local bond-orientational order parameter, $\psi_{6}$ for each particle $i$ is defined as
\begin{equation}
\psi_{6}(r_{ij})={1\over q} \sum_{j=1}^{q} e^{i6\theta(\bf{r_{ij}})},
\end{equation}
where the summation is over all $q$ nearest neighbors  of the particle $i$.  $\theta$ is the angle between the vector $\bf{r}_{ij}$ connecting particle $i$ to $j$ and $x$-axis. In the case of perfect hexagonal symmetry, $|\psi_{6}|=1$ and $\langle|\psi_6|\rangle>0.7$ indicates that the system is crystalline \cite{wu2009melting,nagamanasa2011confined}. 
In our systems, the particles with 5 or 7 neighbors (dislocations) exhibit $\langle|\psi_6|\rangle\approx0.525$, and for an initial configuration, their density is $\approx 5.8\%$  (The rest $\approx 94\%$ particles with hexatic symmetry and display $\langle|\psi_6|\rangle\approx0.985$). Therefore, our study samples display  $\langle|\psi_6|\rangle\gtrsim0.95$. Under deformation, $\langle|\psi_6|\rangle$ even possess the larger value (until  $\gamma_0^\mathrm{TCS}=0.027$ and $\gamma_0^\mathrm{SS}=0.055$) as shown in Figs. \ref{Fig:Fig3}(a) and \ref{Fig:Fig3}(b). The fraction of particles with the largest $|\psi_{6}|$ also increases, as shown in Figs. \ref{Fig:Fig3}(c) and \ref{Fig:Fig3}(d). It indicates that the dislocations, i.e., particles with 5 or 7 neighbors disappear and exhibit perfect hexatic symmetry. The disappearance of dislocations occurs continuously with large fluctuations close to the threshold, $\gamma_{0,t}^\mathrm{TCS(SS)}$, indicated by dashed, blue line in Figs. \ref{Fig:Fig3}(a) and \ref{Fig:Fig3}(b). Similar behavior is observed in the distribution of hexatic order parameters. Further increase in strain amplitude reduces the order parameters, and the fraction of particles with the largest $|\psi_{6}|$ goes down, and the maximum appears at a slightly lower $|\psi_{6}|$ for $\gamma_0^\mathrm{TCS}=0.06$ and $\gamma_0^\mathrm{SS}=0.12$. A similar behavior is observed when shear-induced melting and crystallization were investigated by confocal microscopy in concentrated colloidal suspensions of hard-sphere-like particles \cite{wu2009melting}. A possible reason is that the particles move along a zigzag path \cite{ackerson1990shear,stevens1993simulations,stevens1991shear,derks2009dynamics} and they experience more collisions than without shear. Together with the hydrodynamics under shear, this leads to a larger mean square displacement and thus larger deviations from hexagonal symmetry \cite{derks2009dynamics}. In inset of Fig. \ref{Fig:Fig3}(b), we display $\langle|\psi_6|\rangle$ vs. $\gamma_0^\mathrm{TCS(SS)}-\gamma_{0,t}^\mathrm{TCS(SS)}$ for both modes of deformation. The results show that one can access larger ordering in shear deformation than the tensile-compressive strain. This is connected with the larger diffusivity in TCS mode after the threshold value than in SS mode. To be precise, when we increase the strain amplitude, the number of cycles before achieving reversibility rises. Consequently, there is a greater degree of particle mobility, leading to the annihilation of defects and an increase in the hexatic order parameters. On a passing note, in soft systems like colloidal glasses, this connection between irreversibility transition and structural change is complex and debated. In recent research, it has been shown that the yielding transition of colloidal glass in oscillatory shear can be detected through the static structure of the system \cite{denisov2015sharp}. However, computer simulations differ from the conclusion and argue that yielding is revealed through the dynamic evolution of the system \cite{kawasakimacroscopic}.\\
{\bf{Atomic strain.}} To understand the plastic deformation at the particle level in both modes of deformation, we look at the von Mises shear stress, $\eta$ \cite{Futoshi}. The following algorithm has been used: the initial configuration is considered as reference, and the local transformation matrix $\mathbf{J}_i$ that best maps
$\{\mathbf{d}^0_{ji}\}\rightarrow\{\mathbf{d}_{ji}\}, \forall j \in P_i^0$ is formed, 
where $\mathbf{d}$'s are vector separations (row vectors) between atom $j$ and $i$ (superscript 0 means the reference configuration). Here, $j$ is one of atom $i$'s nearest neighbors, and $P_i^0$ is the total number of nearest neighbors of atom $i$, at the reference configurations. $\mathbf{J}_i$ is determined by minimising $\sum_{j\in P_i^0}|\mathbf{d}_{ji}^0\mathbf{J}_i-\mathbf{d}_{ji}|^2 \rightarrow \mathbf{J}_i$.
For each $\mathbf{J}_i$, the local Lagrangian strain matrix is computed as 
$\eta_i=\frac{1}{2}\left(\mathbf{J}_i \mathbf{J}_i ^T-\mathbf{I}\right)$.
Then, the local shear invariant is calculated for each atom as
\begin{equation}
\eta_i=\sqrt{\eta_{xy}^2+\frac{(\eta_{xx}-\eta_{yy})^2}{2}}.
\end{equation}
and the atomic hydrostatic volumetric strain can be read as 
\begin{equation}
    \delta_i=\frac{\Delta V}{V}\approx\frac{\eta_{xx}+\eta_{yy}}{2}.
    \label{Eq:V_strain}
\end{equation}
We measure the local shear invariant of individual particles after the completion of each cycle, $n$, with respect to the initial configuration, i.e., $n_\mathrm{ref}=0$ and the average, $\langle\eta_{n_\mathrm{ref}=0}^n\rangle$, is plotted as a function of $n$ in Figs. \ref{Fig:Fig4}(a) and \ref{Fig:Fig4}(b), for tensile-compressive and shear strain, respectively. Qualitatively similar behavior is observed, as in the case of particle mobility [see Fig. \ref{Fig:Fig2}]. When $d\langle\eta_{n_\mathrm{ref}=0}^n\rangle/dn$ is measured as a function of $n$ for different values of $\gamma_0^\mathrm{TCS(SS)}$, the observed large transient fluctuations in $d\langle\eta_{n_\mathrm{ref}=0}^n\rangle/dn$ close to the transition indicates the crackling noise, which is a signature of the non-equilibrium transition of grain boundary depinning (data not shown). This echoes the analogous behavior observed in $d\langle\Delta r^2\rangle/dn$ versus $n$ for different $\gamma_0^\mathrm{TCS(SS)}$ (data not shown). Given these findings, we are inclined to adopt a similar methodology as diffusivity--a widely accepted metric for measuring the transition from reversible to irreversible processes. We measured local slope, $d\langle\eta_{n_\mathrm{ref}=0}^n\rangle/dn$, from $\langle\eta_{n_\mathrm{ref}=0}^n\rangle$ vs. $n$ plot and averaged over the steady states, denoted as $\mathcal{F}$, and is shown as a function of strain amplitudes, $\gamma_0^\mathrm{TCS(SS)}$, in Fig. \ref{Fig:Fig4}(c) for TCS and SS. The threshold strain amplitude appears the same as we see in the case of particle mobility. Further analysis reveals that the rate of change of the strain accumulation above the threshold amplitude is larger in the case of TCS than in the case of SS. The increment scales as $\mathcal{F}\sim \left(\gamma_{0}^\mathrm{TCS(SS)}-\gamma_{0,t}^\mathrm{TCS(SS)}\right)^{\beta}$ with $\beta=0.94$ and $1.38$ for TCS and SS, respectively [see Fig. \ref{Fig:Fig4}(d)]. 

To further understand the atomic strain of individual particles, we measure the distribution of $\eta$ after completion of $n_\mathrm{max}(=200)$ cycles with respect to the initial configuration, $P(\eta_{n_\mathrm{ref}=0}^{n_\mathrm{max}})$, and display in Figs. \ref{Fig:Fig5}(a) and \ref{Fig:Fig5}(c) for tensile-compressive and shear strain, respectively. 
Under TCS, when the strain amplitude is kept at a low value, such as $\gamma_0^\mathrm{TCS}=0.021$, approximately $39\%$ of the particles are found within a region where $\eta_{n_\mathrm{ref}=0}^{n_\mathrm{max}}<0.04$, and they are situated in the interior of the grains. Notably, distinct peaks emerge at $\eta_{n_\mathrm{ref}=0}^{n_\mathrm{max}}\approx 0.1$ and at $0.3$, which can be attributed to dislocation motions occurring through the grains (refer to Fig. \ref{Fig:Fig6}(a)). Furthermore, particles experiencing even higher strain are observed because of the dislocation motions near the grain boundaries. Moreover, as the value of $\gamma_0^\mathrm{TCS}$ increases, as anticipated, the occurrence of particles with a strain below $0.04$ becomes less frequent [see Figs. \ref{Fig:Fig6}(b) and \ref{Fig:Fig6}(c)]. Nevertheless, the distinct peaks at approximately 0.1 and 0.3 in the $\eta_{n_\mathrm{ref}=0}^{n_\mathrm{max}}$ distribution persist even with the increase in $\gamma_0^\mathrm{TCS}$. Similar dynamics are evident under shear deformation when $0.03\leq\gamma_0^\mathrm{SS}\approx\gamma_{0,t}^\mathrm{SS}$, as clarified in the strain map in Figs. \ref{Fig:Fig6}(d) and \ref{Fig:Fig6}(e). However, for $\gamma_0^\mathrm{SS}=0.01 \text{ and } 0.02$, no additional peaks are present. Instead, we observe a power-law distribution with an exponent of -1.54. In this range of strain amplitudes, strain accumulation occurs primarily due to the motion of defects near the grain boundaries rather than the motion of the defect through the grains. To be precise, depending on the strain amplitude, we have observed two distinct mechanisms of plastic deformation in polycrystalline systems: dislocations may move through the grains, eventually forming slip lines in two dimensions. Secondly, particles at the grain boundaries undergo local rearrangements, leading to grain boundary motions. These two types of motion have been previously reported in the study by Shiba et al. \cite{shiba2010plastic}. On a passing note, in the case of colloidal glass, the second type of rearrangement (referring to the motion of particles at the grain boundaries) is predominantly observed \cite{spaepen1977microscopic,manning2007strain}.

In Figs. \ref{Fig:Fig5}(b) and \ref{Fig:Fig5}(d), we show the distribution of $p(\eta_{n_\mathrm{ref}=0}^{n_\mathrm{max}}/\langle\eta_{n_\mathrm{ref}=0}^{n_\mathrm{max}}\rangle)$ for TCS and SS, respectively and data collapse reasonably with expected differences before and after the threshold strain amplitudes. For both modes of deformation, when $\gamma_0^\mathrm{TCS(SS)}>\gamma_{0,t}^\mathrm{TCS(SS)}$, the motion of the dislocations through crystals dominates the distribution. However, a key difference is that the strain is accumulated diagonally in TCS [see Fig. \ref{Fig:Fig6}(c)], whereas, in the case of shear loading, the strain accumulation is either along the $x$ axis or along the $y$ axis, as shown in Figs. \ref{Fig:Fig6}(e) and \ref{Fig:Fig6}(f). This observation can be understood based on the Peach-Koehler theory, from where one can write the elastic energy of the slip in isotropic elasticity as 
 \begin{equation}
F_\mathrm{slip}=\frac{Gb^2}{2\pi}\frac{\ln(\mathcal{L}/b)}{1-\nu}\pm\sigma_{xy}^\mathrm{ext}b\mathcal{L}\cos(2\phi).
 \end{equation}
 The first term is the elastic energy of the dislocations, and the second is the work of the applied force. The $\pm$ indicates the direction of motion of the particles around the slip line, $b$ is the lattice constant, $\mathcal{L}$ is the slip length, $G$ is the shear modulus and $\nu$ is the Poisson’s ratio and the $\phi$ is the angle between the slip direction and the $x$-axis. For shear deformation with $\sigma_{xy}^\mathrm{ext}>0$ and when we use $+$, one can achieve the lowest $F_\mathrm{slip}$ for $\phi=0$ (i.e., the slip is along $x$ axis), and when we use $-$, the slip is along $y$ axis as $\phi=\pi/2$. For uni-axial deformation $\sigma_\alpha^\mathrm{ext}$, one can write an analogous equation as follows:
 \begin{equation}
F_\mathrm{slip}=\frac{Gb^2}{2\pi}\frac{\ln(\mathcal{L}/b)}{1-\nu}\pm\sigma_{\alpha}^\mathrm{ext}b\mathcal{L}\sin(2\phi)
 \end{equation}
and the preferred orientation of the slip-line is $\pm\pi/4$. These preferred directions have been observed in amorphous metals \cite{schuh2007mechanical,wei2002evolution}  and granular materials \cite{desrues1985localization,desrues2002shear}. In simulations, shear bands in these preferred directions have been realized in model amorphous metals and polymers \cite{deng1989simulation,shi2006atomic,maloney2008evolution,bailey2006atomistic,bulatov1994stochastic,argon1995plastic} and in a model crystal with weak elastic anisotropy \cite{onuki2003plastic}.
\begin{figure}
    \centering
\includegraphics[scale=0.45]{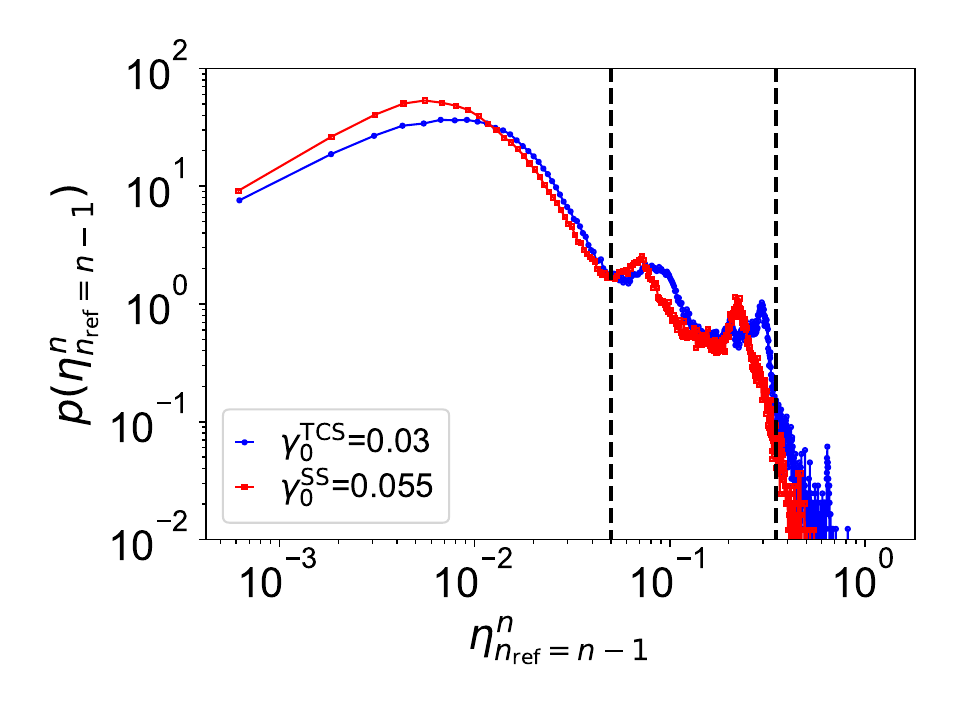}
   \caption{von-Mises shear strain of individual particle is measured after each cycle, $n$, with respect to the configuration after $n-1$ cycle, and the distribution averaged over the first 20 cycles displayed. For both tensile-compressive and shear strains, two peaks appear between $0.05$ and $0.35$.}
    \label{Fig:Fig8}
\end{figure}
\begin{figure*}
    \centering
\includegraphics[scale=0.5]{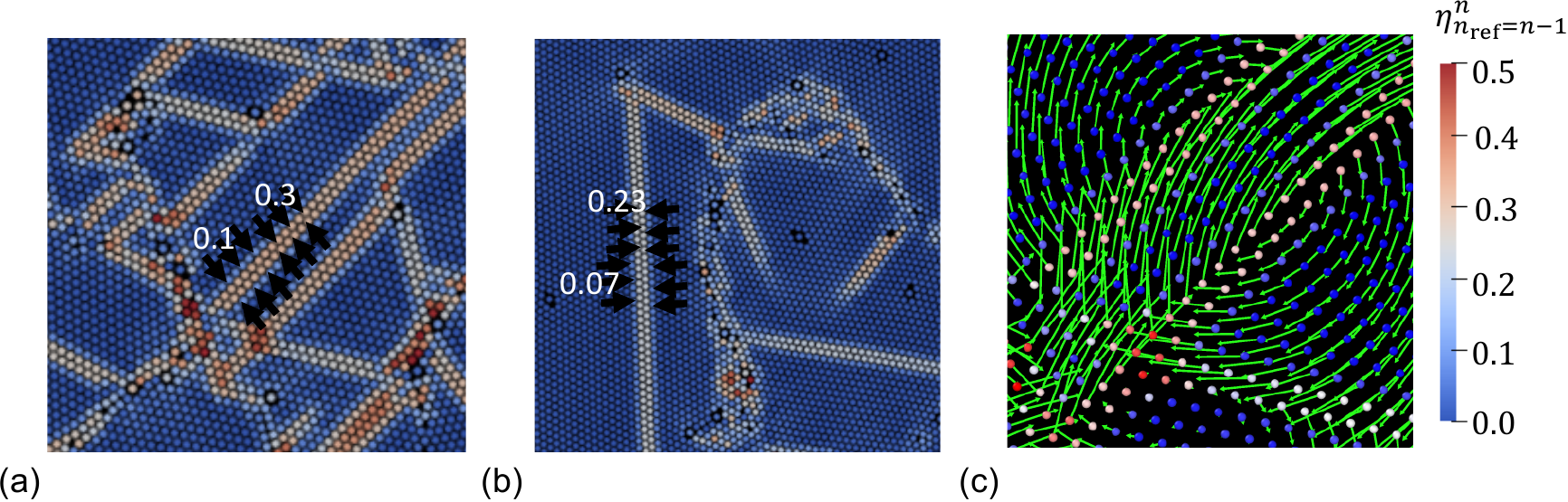}
   \caption{von-Mises shear strain map for individual particles measured with respect to the previous configuration to display the dislocation motions through crystals. (a) is for  $\gamma_0^\mathrm{TCS}=0.03$ and (b) is for $\gamma_0^\mathrm{SS}=0.055$. (c) Displacement map corresponding to the motion of the defect in oscillatory TCS. The arrow indicates the direction of the displacement.}
\label{Fig:Fig9}
\end{figure*} 
\begin{figure*}
    \centering
        (a)\includegraphics[scale=0.4]{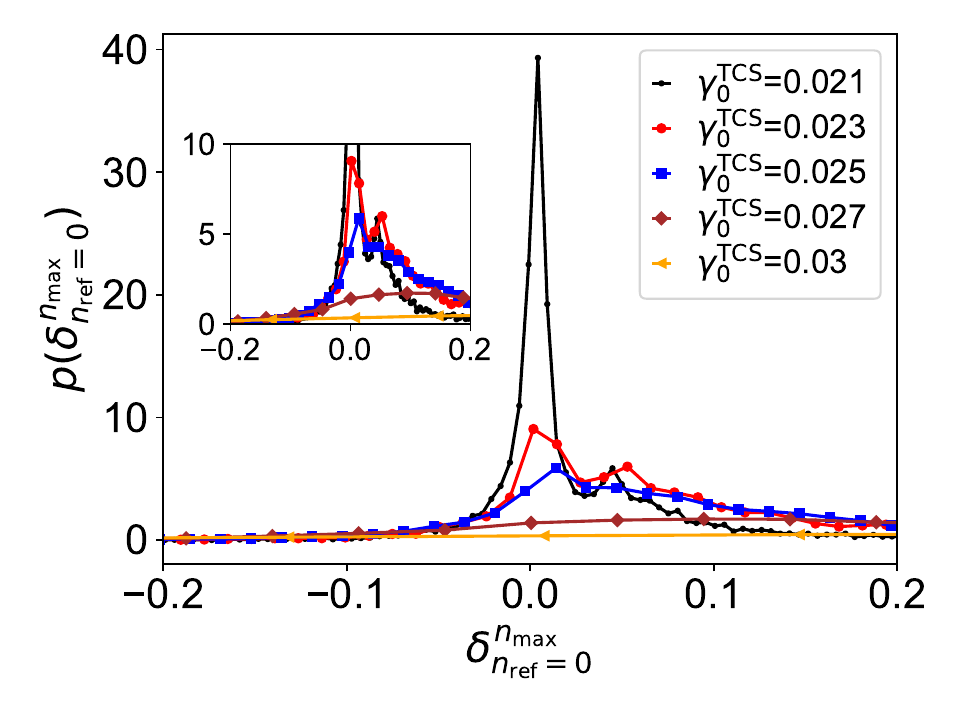}
        (b)\includegraphics[scale=0.4]{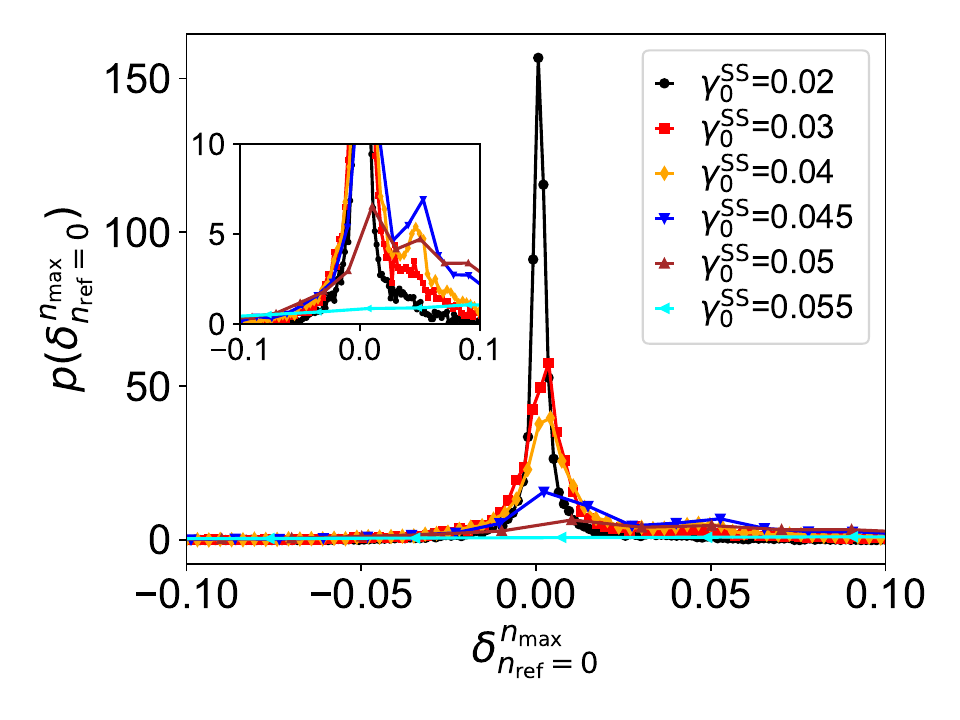}
        (c)\includegraphics[scale=0.4]{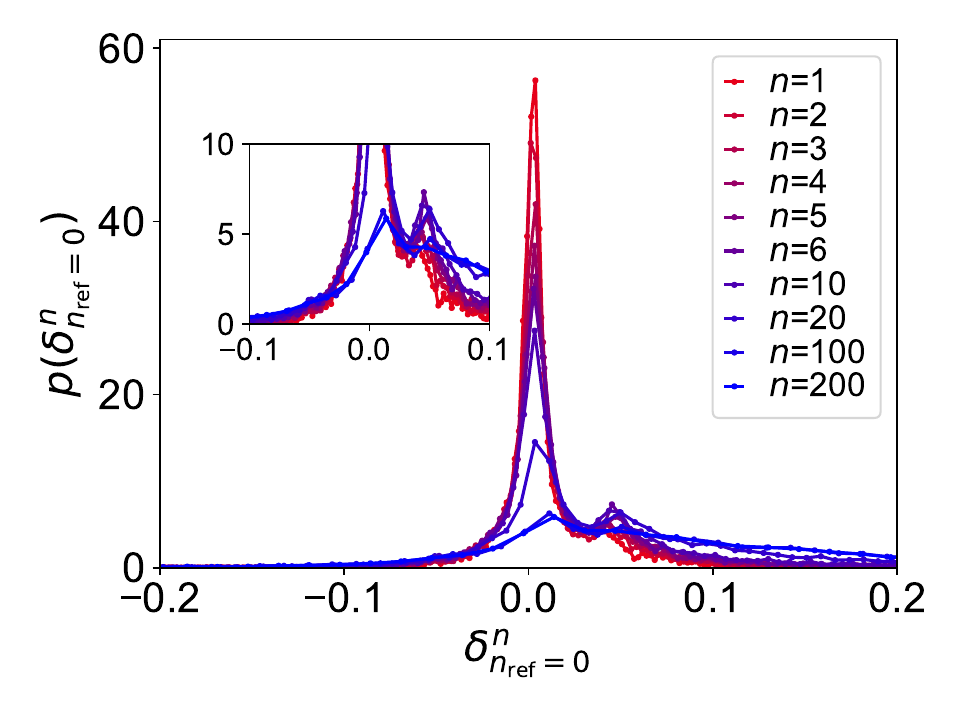}
        (d)\includegraphics[scale=0.4]{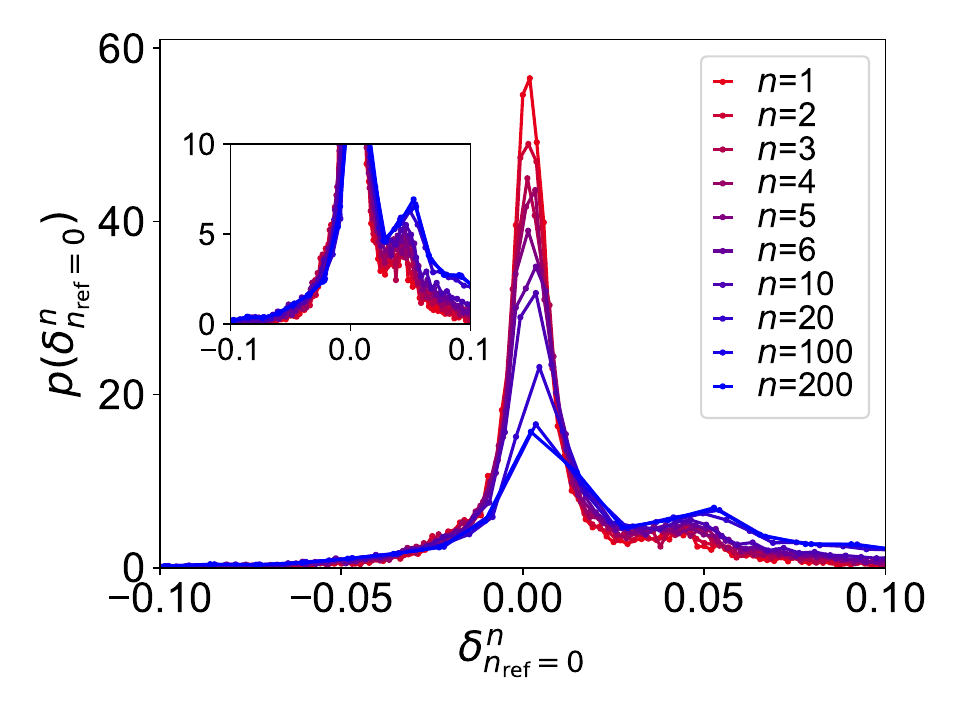}
   \caption{(a) Distribution of volumetric strain of individual particle, $\delta_i$, measured after $n_\mathrm{max}(=200)$ cycles for different strain amplitudes of (a) tensile-compressive and (b) shear loading. The distribution is measured after each cycle, $n$, and shown as mentioned in the legend of (c) and (d). (c) is for tensile-compressive strain with $\gamma_0^\mathrm{TCS}=0.025$ and (d) is for shear strain with $\gamma_0^\mathrm{SS}=0.045$. The inset in the figures is zoomed in on the additional peak.}
    \label{Fig:Fig10}
\end{figure*}

To further analyze the atomic strain, we compute $p(\eta_{n_\mathrm{ref}=0}^{n})$ after each $n$ for $\gamma_0^\mathrm{TCS}=0.025$ [see Fig. \ref{Fig:Fig7}(a)] and $\gamma_0^\mathrm{SS}=0.045$ [see Fig. \ref{Fig:Fig7}(b)] that are close to $\gamma_{0,t}^\mathrm{TCS(SS)}$ and in the reversible regime. Under TCS ($\gamma_0^\mathrm{TCS}=0.025$), we observe that rearrangements around the grain boundaries and dislocation motions through the grains are activated immediately after the first cycle. However, it is worth noting that more than $50\%$ of the particles (approximately $58\%$) remain in the strain regime where the strain $\langle\eta_{n_\mathrm{ref}=0}^{n}\rangle$ is less than $0.04$. 
As the number of cycles $n$ increases, the situation evolves. At $n=30$, we observe indications of two additional peaks in the probability distribution function $p(\eta_{n_\mathrm{ref}=0}^{n})$. However, these peaks become smeared out when $n=200$. On the other hand, in the case of shear strain ($\gamma_0^\mathrm{SS}=0.045$), the two additional peaks remain clearly visible in the strain regime under investigation even at $n=200$.  However, a situation similar to that for tensile-compressive deformation (i.e., peaks become smeared out when $n=200$) emerges for $\gamma_0^\mathrm{SS}=0.05$. We observe a resemblance in the behavior of atomic rearrangements between the two cases, indicating the common characteristics in the strain response under both shear and tensile-compressive strain.

To identify and characterize the two additional peaks in both modes of deformation, we compute the von Mises shear strain for individual particles at the cycle number $n$ with respect to the previous cycle, $n-1$, instead of taking the configuration where $n=0$ as reference. Distribution and the strain map are shown in Fig. \ref{Fig:Fig8} and in Fig. \ref{Fig:Fig9}, respectively, for $\gamma_0^\mathrm{TCS}=0.03$ and $\gamma_0^\mathrm{SS}=0.055$ which are in the irreversible regime where the defect motion through crystal dominates. Two clear peaks are observed in both cases [see Figs. \ref{Fig:Fig8}, \ref{Fig:Fig9}(a), and \ref{Fig:Fig9}(b)]. When dislocations move through the crystals, the particles at the adjacent grains exhibit either clockwise or anti-clockwise rotation. See the displacement field in Fig. \ref{Fig:Fig9}(c). This type of rotation leads to the merging of the grains with different orientations and creates a single-grain structure. 

Lastly, we evaluate the volumetric strain ($\delta$) of each particle using Eq. \ref{Eq:V_strain}, and their distributions are displayed in Figs. \ref{Fig:Fig10}(a) and \ref{Fig:Fig10}(b) for different values of TCS and SS amplitudes, respectively. When the shear strain amplitude is small [see $\gamma_0^\mathrm{SS}=0.02$ in Fig. \ref{Fig:Fig10}(b)], we observe a symmetrical distribution centered around $\delta=0$. However, with the increase of $\gamma_0^\mathrm{SS}$ (say 0.04, 0.045), the distribution becomes positively skewed, and the additional peak appears and grows around $0.05\pm0.01$ [see the inset of Fig. \ref{Fig:Fig10}(b)] and the peak height decreases when the strain amplitude approaches the threshold value. This additional peak in volumetric strain is also the signature of dislocation motion through the grains. This motion through grains takes over the volumetric strain distribution in the irreversible regime. Under TCS, the system behaves similarly [see Fig. \ref{Fig:Fig10}(a) and the inset]. The evolution pattern of the additional peak over the cycles at a strain amplitude close to the threshold value appears non-monotonic depending on the strain amplitude. We see the nonmonotonicity when $\gamma_0^\mathrm{TCS}=0.025$, as shown in Fig. \ref{Fig:Fig10}(c) and $\gamma_0^\mathrm{SS}=0.05$ (data not shown) while it is not at $\gamma_0^\mathrm{SS}=0.045$ [see Fig. \ref{Fig:Fig10}(d)]. In addition to that, we also observe a positive skewness in the distribution as we apply more and more cycles. Additionally, we observe a significant correlation between the volumetric strain and the shear strain in both modes of deformation. This finding is consistent with previous observations in the case of metallic glasses \cite{wang2016direct}, indicating a common behavior across different materials and deformation modes. \section{\label{sec:level1}Conclusions}
The present study investigates the irreversible transition in a 2D polycrystalline sample under two distinct oscillatory deformation modes: tensile-compressive and shear strain. Upon subjecting the material to both deformation modes, we observe a reversible elastic response up to a critical strain amplitude, beyond which irreversible plastic deformation occurs. Notably, the threshold strain amplitude required for the transition is found to be larger in the case of shear strain compared to the TCS. To confirm the threshold values of the reversible-irreversible transition, we conduct dynamical analyses, including assessments of particle mobility and atomic strain. Additionally, we perform structural analyses, such as evaluating the hexatic order parameter. 
In particle mobility, we have studied the MSD of the particles with respect to the initial configurations. From there, we define diffusivity as the local slope of the MSD vs. the number of cycles plot. We found that for both modes of deformation, the diffusivity decreases $\approx 1 \text{ to } 2$ order of magnitudes close to the threshold value from the above, defined as the irreversible regime. Then jump to the value which is $\mathcal{D}\approx0$ during the steady states, defined as the reversible regime. On a passing note, one may consider the average particle displacement after each cycle as a dynamic observable instead of the diffusivity from the MSD as done in Ref. \cite{kawasakimacroscopic}. This study also opens up the question of whether the non-equilibrium transition is continuous or discontinuous.

In the structural analysis, we have observed that the hexatic order parameter increases as we increase the strain amplitude for both modes of deformation, which indicates the disappearance of grain boundaries and dislocations. Further increase in strain amplitudes lowers the order parameters.

To explore the strain at the particle level, we measure both von Mises shear strain and the volumetric strain. The change of von Mises shear strain per cycle as a function of strain amplitude shows qualitatively similar behavior as diffusivity for both modes of deformation. We observe two types of dynamics: first, the local rearrangement of particles near the grain boundary leads to the grain boundary motion, and second, the motion of dislocation through the grains results in a slip line in two dimensions. Further analysis reveals that in the case of tensile-compressive shear, the strain is accumulated (slip line) diagonally, whereas, for shear strain, the strain accumulation occurs along the $x$ and the $y$ directions. The nature of the strain accumulation is explained in terms of the Peach-Koheler theory. The volumetric strain analysis also captures the dislocation motion through the grains. Additionally, we observe a strong correlation between the von-Miese shear strain and the volumetric strain. 

Our findings reveal that colloidal polycrystals subjected to oscillatory shear and tensile-compressive strain exhibit irreversibility transition similar to disordered particle assemblies and colloidal glasses. However, in this context, the relevant degrees of freedom are the topological defects rather than the particles themselves. The implications of our study extend to applications such as Zener pinning, where inclusions are strategically utilized to impede grain growth in polycrystals \cite{hazzledine1990computer}. The critical strain amplitude for irreversibility that we identified should correspond to the depinning stress required to initiate grain boundary growth \cite{moretti2004depinning}. One can test our results in colloidal particle experiments 
\cite{Ghofraniha,Louhichi,Keim}.
This study can potentially expand into the realm of amorphous solids \cite{Regev,parmar2019strain}. Our investigation revealed that the larger strain amplitude of irreversible transition in SS mode is linked to a deformation geometry. Based on this finding, we anticipate a similar effect in amorphous solids when subjected to these deformation modes.
\section{\label{sec:level1}Acknowledements} 
P.K.J.
acknowledges the financial support from the Science and Engineering
Research Board, Government of India (SRG/2022/000993) and ACRG from BITS Pilani. Khushika acknowledges the financial support from the BITS Pilani, Pilani Campus Institute fellowship. 
The simulations
were performed at the high-performance computing cluster of BITS Pilani, Pilani Campus. 
\bibliography{literature}
\end{document}